\documentclass[useAMS,usenatbib,a4paper]{mn2e}
\usepackage{graphicx}
\usepackage{amsmath}
\usepackage{amssymb}

\title[On the exclusion of intra-cluster plasma from AGN-blown
  bubbles]{On the exclusion of intra-cluster plasma from AGN-blown
  bubbles}

\author[E.C.D. Pope] {Edward
  C.D. Pope$^{1}$\thanks{E-mail:ecdpope@uvic.ca}\\$^{1}$Department of
  Physics \& Astronomy, University of Victoria, Victoria, BC, V8P 1A1,
  Canada\\}

\begin{document}

\pagerange{\pageref{firstpage}--\pageref{lastpage} \pubyear{2009}}

\maketitle

\label{firstpage}

\begin{abstract}
Simple arguments suggest that magnetic fields should be aligned
tangentially to the surface of an AGN-blown bubble. If this is the
case, charged particles from the fully ionised intra-cluster medium
(ICM) will be prevented, ordinarily, from crossing the boundary by the
Lorentz force. However, recent observations indicate that thermal
material may occupy up to 50\% of the volume of some bubbles. Given
the effect of the Lorentz force, the thermal content must then be
attributed to one, or a combination, of the following processes: i)
the entrainment of thermal gas into the AGN outflow that inflated the
bubble; ii) rapid diffusion across the magnetic field lines at the
ICM/bubble interface; iii) magnetic reconnection events which transfer
thermal material across the ICM/bubble boundary. Unless the AGN
outflow behaves as a magnetic tower jet, entrainment may be
significant and could explain the observed thermal content of
bubbles. Alternatively, the cross-field diffusion coefficient required
for the ICM to fill a typical bubble is $\sim 10^{16}\,{\rm
  cm^{2}\,s^{-1}}$, which is anomalously high compared to predictions
from turbulent diffusion models. Finally, the mass transfer rate due
to magnetic reconnection is uncertain, but significant for plausible
reconnection rates. We conclude that entrainment into the outflow and
mass transfer due to magnetic reconnection events are probably the
most significant sources of thermal content in AGN-blown bubbles.
\end{abstract}


\section{Introduction}
Observational studies indicate that powerful radio emission from
Active Galactic Nuclei (AGN) at the centres of galaxy clusters is
strongly related to the thermal state of the hot, X-ray emitting
plasma that permeates the cluster. In particular, systems with short
radiative cooling times at the cluster centre seem more likely to
exhibit on-going star formation, optical line-emission and radio
emission from AGN
\citep[e.g.][]{burns,crawf,raff08,cav08,mittal}. There is also a
significant body of theoretical evidence indicating that outflows from
AGN play an important role in the evolution of their surroundings
\citep[e.g.][]{bintab,benson,croton05,bower06,short,puchwein,bower08,pope09}. As
a result, it is often assumed that AGN operate as thermostats which
balance and regulate the radiative losses of the intra-cluster medium
(ICM) \citep[e.g.][]{stats}. In the standard paradigm, material cools
out of the ICM and is accreted by a supermassive black hole located in
the central galaxy. This releases vast amounts of energy, often in the
form of outflows, which couple to and heat the ambient gas, thereby
reducing the cooling rate of the ICM \citep[e.g.][]{chur06}.

To understand how the energy of the outflow is actually dissipated and
heats the ICM, it is necessary to build an accurate model describing
the interaction between the outflow and the ICM. In this article, we
focus specifically on the interaction between the ICM and the
bubble-like structures inflated by AGN outflows. Observations indicate
that these bubbles are largely devoid of thermal material
\citep[e.g.][]{mcnuls} and primarily consist of magnetic fields and
high-energy particles \citep[c.f.][]{dunn04,birz08}. In order for the
bubbles to appear as depressions in the X-ray surface brightness
\citep[e.g.][]{mcnuls}, we infer that the mass of thermal material
entering from the ICM must be small compared to the mass initially
displaced during the inflation. Therefore, the observations seem to
indicate that the bubbles do not mix significantly with their
surroundings. In addition, \cite{sand} argue, based on an analysis of
the bubbles in the Perseus cluster, that the volume fraction of
thermal material in bubbles is no more than 50\%.

It is generally supposed that mixing is mostly inhibited by the
presence of magnetic fields at the interface between the bubble and
the ICM. In particular, following the work of \cite{chand},
\cite{deyoung03,instab105,dursi07} showed that a magnetic field
oriented tangentially to the bubble surface may prevent the growth of
Rayleigh-Taylor instabilities. This has also been shown numerically by
\cite{robinson04,on}. However, see also \cite{pizz2,scanna} for other
explanations of bubble stability.

One possibility for the origin of such a favourable magnetic field
orientation is that, during the expansion of the bubble, the ambient
medium, along with its magnetic field, is excluded from the
bubble. This results in a sheath around the bubble in which the field
is primarily tangential to the bubble surface, and which has a higher
value than the ambient value, due to the effects of compression
\citep[][]{deyoung03}. Numerical simulations of moving bubbles
\citep[see][]{rusz07, dursi} have also shown that that magnetic field
from the ICM can become draped across the front of the bubble. This
ensures that, whatever the initial configuration of the magnetic field
on the bubble surface, there should always be layer of magnetic field
that is oriented tangentially to the bubble surface.

However, suppressing the growth of fluid instabilities on the bubble
surface is only one aspect of the interaction between the ICM and the
bubble. That is, the magnetic field on the bubble surface might
provide a stable framework which prevents the bubble from fragmenting
into many smaller bubbles, but this does not necessarily prevent
thermal material entering the bubble. An important and related effect
arises from the fact that the ICM is a fully ionised plasma and
therefore has a high electric conductivity, while the AGN-blown bubble
contains a magnetic field. This scenario is extremely similar to the
interaction between the solar wind and the Earth's magnetosphere
\citep[see for example][]{chap,kivelson, parks}. Therefore, in the
idealised case, where the ICM carries no magnetic field, we expect a
charge entering the bubble to experience a Lorentz force which is
perpendicular to both its direction of motion and the magnetic
field. Provided the bubble magnetic field is oriented tangentially to
the bubble surface, the charge will re-directed out of the bubble and
emerge back in the ICM. The combination of this exclusion and the
stabilising effect of magnetic fields creates a relatively impermeable
barrier between the bubble and the ICM. Furthermore, the ICM is not a
perfect conductor and some of the electromagnetic energy, generated by
the interaction, will be dissipated as heat in the region around the
bubble. In principle, this could heat the ICM around the bubble, or
help to power the optical emission from cool material observed behind
some bubbles.

The plasma exclusion effect described above has significant
implications for interpreting observations of bubbles. In particular,
any thermal material could only have entered the bubble by the
following processes: i) thermal material entrained into the outflow
that originally inflated the bubble; ii) particle diffusion across the
magnetic field on the bubble surface; iii) non-ideal processes such as
magnetic reconnection, which occur as a result of non-zero resistivity
in the ICM and the interaction between magnetic fields in the ICM and
the bubble. With reference to the first process, mass-loading will
decelerate the AGN outflow \citep{hart86} and so may potentially
explain the absence of strong shocks in the centres of galaxy clusters
\citep[e.g.][]{form8707,grah}. The velocity of the outflow also plays
a role in governing the morphology of the subsequent structure. For
example, slower outflows are more likely to inflate bubble-like
structures, rather than the elongated structures produced by faster
jets \citep[e.g.][]{stern08}. Therefore, determining the origin of a
bubble's thermal content is of importance in the wider context of AGN
feedback.

The article is structured as follows: in section 2, we introduce the
model which describes the interaction between the ICM and the bubble
magnetic field. The main quantities derived from this model, such as
the Ohmic heating rate and induced current are given in section
3. Section 4 provides a discussion of the mass transport rates
associated with outflow entrainment, diffusion and magnetic
reconnection. We summarise the findings in section 5.

\section{Model}

In this derivation, the ICM is taken to be an unmagnetised, fully
ionised plasma consisting only of electrons and protons. The bubble's
magnetic field is taken to be aligned tangentially to the bubble
surface, see figure 1. We also assume that both the ICM and bubble
fluids can be described using the equations of ideal
magnetohydrodynamics. This assumption is valid in the limit that the
diffusion of magnetic fields is much less important than their
advection. The relative importance of diffusion and advection for
magnetic fields is often quantified using the magnetic Reynolds
number, $Re_{\rm m} = UL/\eta $, where $U$ and $L$ are the velocity
and length scales of interest and $\eta$ is the electric
resistivity. For a fully ionised plasma $\eta$ is given by
\citep[e.g.][]{boden}
\begin{equation}\label{eq:1}
\eta \approx 260 \bigg(\frac{\ln \Lambda}{40}\bigg)
\bigg(\frac{T}{10^{8}\,{\rm K}}\bigg)^{-3/2}\,{\rm cm^{2}\,s^{-1}}.
\end{equation}
where $\ln \Lambda$ is the Coulomb logarithm. Therefore,
\begin{equation}\label{eq:2}
Re_{\rm m} \approx 6 \times 10^{27} \bigg(\frac{U}{10^{8}\,{\rm
    cm\,s^{-1}}}\bigg)\bigg(\frac{L}{5\,{\rm
    kpc}}\bigg)\bigg(\frac{T}{10^{8}\,{\rm K}}\bigg)^{3/2}.
\end{equation}
Since $Re_{\rm m} \gg 1$, the diffusion of magnetic fields is
unimportant on the length scale of interest, $L$. More specifically,
we can say that a magnetic field initially residing in an AGN-blown
bubble cannot diffuse far and so is effectively frozen to the bubble
fluid. This is the so-called `frozen-in' approximation, which
corresponds to the flow regime known as ideal magnetohydrodynamics.

The interaction between the ICM and the magnetised AGN-blown bubble
can be understood by using a combination of fluid and particle
approaches. The fluid model provides a description of the conditions
at the ICM/bubble interface, showing that in order to prevent thermal
ICM plasma entering the bubble there must be an electric current
flowing in the boundary. The particle description shows that the ICM
is excluded from the majority of the bubble due to the Lorentz force
acting on the charges, but that the partial penetration and deflection
of these particles is sufficient to generate the current. These
methods are described in detail below.

\subsection{The fluid model}

Observationally, AGN-blown bubbles appear as depressions in the X-ray
surface brightness \citep[e.g.][]{mcnuls}, meaning that they contain
comparatively little material emitting at these energies. Recent
evidence of bubbles in the Perseus cluster \citep[][]{sand} suggests
that the volume fraction of thermal material is less than 50\% meaning
that the fluid pressure across the boundary must be discontinuous
\citep[e.g.][]{parks}. The conditions at the boundary must satisfy the
fluid momentum equation. For a species, $i$, of charged fluid, the
momentum equation is
\begin{equation}\label{eq:3}
\rho_{\rm i} \frac{\partial \mathbf{v_{\rm i}}}{\partial t} +
\rho_{\rm i}(\mathbf{v_{\rm i}}.\mathbf{\nabla})\mathbf{v_{\rm i}} =
-\mathbf{\nabla} P_{\rm i} + \mathbf{J_{\rm i}} \times \mathbf{B} +
\rho_{\rm i} \mathbf{F},
\end{equation}
where $\rho_{\rm i}$ is the density of the fluid with pressure,
$P_{\rm i}$, $\mathbf{J_{\rm i}}$ is the current density vector,
$\mathbf{B}$ is the magnetic field vector and $\mathbf{F}$ is an
external force per unit mass, such as gravity. The $\mathbf{J_{\rm i}}
\times \mathbf{B}$ term describes the reaction of the fluid to the
magnetic field.

Therefore, assuming that the external force, $\mathbf{F}$, is locally
unimportant, the equilibrium condition to be satisfied at the bubble
surface is $\mathbf{\nabla} P_{\rm i} = \mathbf{J_{\rm i}} \times
\mathbf{B}$. The magnitude and direction of the induced current must
then be \citep[e.g.][]{parks}
\begin{equation}\label{eq:4}
\mathbf{J_{\rm i}} = \frac{\mathbf{B}\times \mathbf{\nabla} P_{\rm
    i}}{B^{2}}.
\end{equation}
Furthermore, the current calculated from equation (\ref{eq:4}) must
also satisfy Amp\`{e}re's law
\begin{equation}\label{eq:5}
\mathbf{\nabla} \times \mathbf{B} = \mu_0 \mathbf{J} + \mu_0
\epsilon_0 \frac{\partial \mathbf{E}}{\partial t},
\end{equation}
where $\mu_0$ is the vacuum permeability, and $\epsilon_0 \partial
\mathbf{E} /\partial t$ is known as the displacement current, which
can be neglected. This must produce the $\mathbf{J} \times \mathbf{B}$
force needed to balance the rate of change of ICM momentum. In the
simplest case, where there is no coupling of energy or momentum across
the boundary, these currents close on themselves.

An estimate of the induced current density, $\mathbf{J}$, requires a
value for either $\mathbf{\nabla} P_{\rm i}$ or $\mathbf{\nabla}
\times \mathbf{B}$ which can be calculated using the particle model,
described below.

\subsection{Particle model}

In the scenario depicted in figure 1, the incident ICM particles move
in the negative $x$-direction with a velocity of magnitude $u$. The
magnetic field points in the $z$-direction, with a magnitude $B$ so
that the Lorentz force acting on an incident proton is in the
$y$-direction, with a magnitude $|\mathbf{F}|= euB$. Once moving in
the $y$-direction, the Lorentz force on the proton is then directed in
the $x$-direction causing it to be excluded from the bubble. The
transverse velocity component (in the $y$-direction) that arises
during the half-orbit generates an electric current in the boundary
which prevents the magnetic field from entering the ICM, and is
usually referred to as the Chapman-Ferraro current
\citep[e.g.][]{chap,funaki}. Therefore, once back in the ICM, a charge
experiences no further deflection.

\begin{figure*}
\centering
\includegraphics[width=12cm]{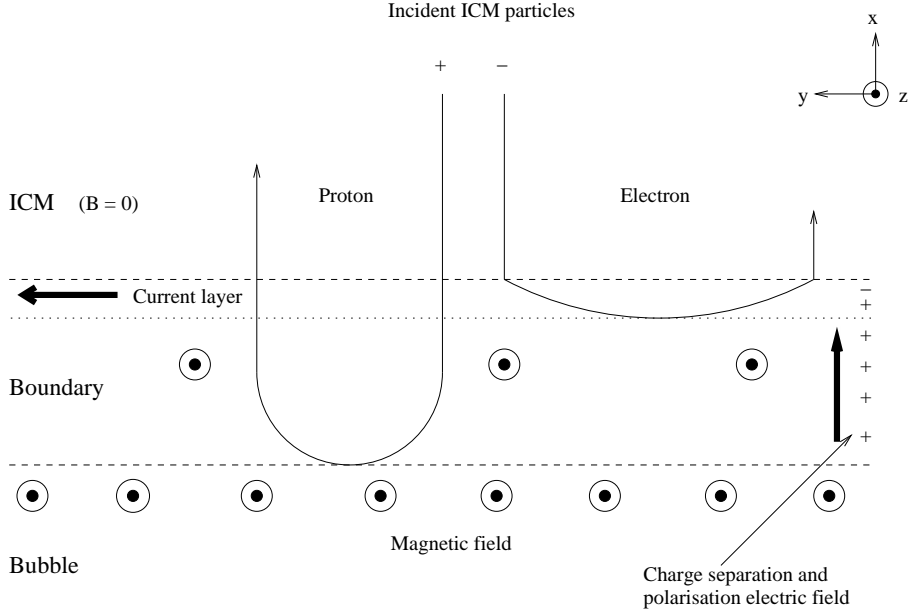}
\caption{Schematic showing the interaction of the ICM with the
  magnetic AGN-blown bubble, based on Baumjohann \& Treumann (1996)
  and Parks (2003).}
\label{fig:flow}
\end{figure*}

For the idealised case, the thickness of the interaction boundary
would be approximately the plasma skin depth, $\delta = c/\omega_{\rm
  p}$, where $c$ is the speed of light and $\omega_{\rm p}$ is the
plasma frequency \citep[][]{funaki}. However, because of their heavier
mass, protons tend to penetrate more deeply into the magnetic field
than electrons. This leads to charge separation with a negative charge
layer above the positive charge layer. Thus, the outwardly pointing
electric field restrains the protons, while the electrons gain energy
\citep[e.g.][]{baumjohann}.

The thickness of the boundary is therefore considered to be larger
than $\delta$, being roughly the proton gyration radius
\begin{equation}\label{eq:6}
r_{\rm g,p} = \frac{m_{\rm p} v_{\rm p}}{e B}
\end{equation}
where $m_{\rm p}$ is the proton mass, $v_{\rm p}$ is the velocity
perpendicular to the field line, $e$ is the electric charge, and $B$
is the magnetic field strength. Since the bubble ascends subsonically,
the particle velocity is largely governed by its thermal motion. Then,
for a thermal proton, the mean velocity is given by $v_{\rm p} = (3
k_{\rm B}T/m_{\rm p})^{1/2}$, where $k_{\rm B}$ is the Boltzmann
constant and $T$ is the temperature. Therefore, assuming that the
electrons and protons have the same temperature, the ratio of the
thermal proton and electron gyration radii is $(m_{\rm p}/m_{\rm
  e})^{1/2}$.

The total current density in the $y$-direction is the sum of the
individual currents generated by the protons and electrons:
$|\mathbf{J}| = \displaystyle\sum_{i=p,e}J_{\rm i} = (n_{\rm p}e
v_{\rm p} + n_{\rm e}ev_{\rm e})$, where $n_{\rm p}$ and $n_{\rm e}$
are the number density of protons and electrons, respectively. $v_{\rm
  p}$ and $v_{\rm e}$ are the proton and electron velocities.

Then, using Amp\`{e}re's law, the magnitude of the induced current
density is roughly
\begin{equation}\label{eq:9}
\displaystyle\sum_{i=p,e}|J_{\rm i}| \approx
\displaystyle\sum_{i=p,e}\frac{B}{\mu_0 r_{\rm g,i}},
\end{equation}
where $r_{\rm g,i}$ is the gyration radius of the $ith$ particle
species and $P_{\rm i}$ is the pressure of the $i$th species.

We also note that the current flows so as to exclude the bubble's
magnetic field from the ICM, which increases the field strength just
inside the boundary region. The numerical value of the field inside
the boundary is then twice the initial value. As a result, the current
layer confines the bubble's magnetic field to the region inside the
boundary and so partly governs the size of the bubble.

\section{Results}

Since the electric conductivity of the ICM is finite, some of the
electromagnetic energy generated by the interaction will be dissipated
as heat in the region around the bubble. In this section, we present a
calculation of the expected heating rate. The magnitude of the induced
current is also of interest, primarily for comparison with the values
required in models of current-dominated jets
\citep[e.g.][]{nakamura}. Accordingly, a calculation for the total
current flowing across the surface of a bubble is also given in this
section.

\subsection{Heating rate}

From equations (\ref{eq:6}) and (\ref{eq:9}), it is clear that
particles with smaller gyration radii contribute a larger current
density. Because of this effect, the induced current density due to
electrons is a factor of $(m_{\rm p}/m_{\rm e})^{1/2}$ greater than
for protons and has a magnitude
\begin{equation}\label{eq:11}
|J_{\rm e}|\approx 4 \times 10^{-5}\bigg(\frac{B}{10\,{\rm \mu
    G}}\bigg)^{2}\bigg(\frac{T}{10^{8}\,{\rm K}}\bigg)^{-1/2}\,{\rm
  A\,cm^{-2}}.
\end{equation}
Although the electric conductivity of the ICM is extremely high, it is
not infinite. Consequently, the induced current dissipates
electromagnetic energy at a rate $\dot{e} = \eta|J|^{2}$, where $\eta$
is the electric resistivity given by equation (\ref{eq:1}). This
process is known as Joule or Ohmic dissipation. Due to the $J^{2}$
term, the Ohmic dissipation rate is a factor $(m_{\rm p}/m_{\rm e})$
higher for electrons than protons and can be written
\begin{equation}\label{eq:13}
\dot{e} \approx \eta|J_{\rm e}|^{2} \approx 3 \times
10^{-18}\bigg(\frac{B}{10\,{\rm \mu
    G}}\bigg)^{4}\bigg(\frac{T}{10^{8}\,{\rm K}}\bigg)^{-5/2}\,{\rm
  erg\,s^{-1}\, cm^{-3}}.
\end{equation}
A plasma with a temperature of $T \sim 10^{8}\,{\rm K}$ and a number
density $n \sim 0.01\,{\rm cm^{-3}}$ primarily radiates energy by
thermal bremsstralung at a rate $n^{2}\Lambda(T) \sim 10^{-27}\,{\rm
  erg\,s^{-1}\, cm^{-3}}$. As can be seen, from equation (\ref{eq:13})
the Ohmic dissipation rate within the boundary region is a factor of
$\sim 10^{9}$ greater than the radiative losses due to thermal
bremsstrahlung in the ICM surrounding the bubble. This is surprisingly
large, suggesting that Ohmic dissipation might have an observable
effect. To determine whether this is true it is necessary to calculate
the total rate at which energy is dissipated.

Assuming that the bubble is spherical, the volume within which the
electromagnetic energy is dissipated is $ (4/3) \pi [(r_{\rm b} +
  r_{\rm g})^{3} - r_{\rm b}^{3}] \approx 4 \pi r_{\rm b}^{2} r_{\rm
  g}$. Because of the extra factor of $r_{\rm g}$, the
volume-integrated dissipation rate for electrons is a factor $(m_{\rm
  p}/m_{\rm e})^{1/2}$ greater than for protons, so electron
dissipation still dominates the total Ohmic dissipation rate
\begin{eqnarray}\label{eq:14}
\dot{E} \approx 4\pi r_{\rm b}^{2}r_{\rm g,e}\dot{e} \approx 2 \times
10^{35}\bigg(\frac{r_{\rm b}}{5\,{\rm kpc}}\bigg)^{2}
\bigg(\frac{B}{10\,{\rm \mu G}}\bigg)^{3}\\ \nonumber \times
\bigg(\frac{T}{10^{8}\,{\rm K}}\bigg)^{-2}\,{\rm erg\,s^{-1}}.
\end{eqnarray}
It is also instructive to demonstrate how the heating rate may vary as
the bubble evolves. In a stratified atmosphere, the rising, buoyant
bubble expands to maintain pressure equilibrium with its surroundings.
For the present case, pressure equilibrium requires that
$B^{2}/(2\mu_0) = P$, where $P$ is the ambient pressure. Then,
assuming the expansion is adiabatic, the bubble volume, $V_{\rm b}$,
is related to the ambient pressure, $P$, by $P V_{\rm b}^{\Gamma} =
P_{\rm b,0} V_{\rm b,0}^{\Gamma}$, where $P_{\rm b,0}$ is the ambient
pressure where the bubble was inflated, $V_{\rm b,0}$ the initial
bubble volume and $\Gamma$ is the adiabatic index of the bubble. A
magnetically-dominated bubble can be described by a relativistic
equation of state, for which $\Gamma = 4/3$. Finally, the ICM pressure
can be approximated by a $\beta$-model: $P = P_{\rm 0} [1+
  (z/z_0)^{2}]^{-\beta}$, where $P_{\rm 0}$ is the central pressure,
$z$ is the distance from the cluster centre, $z_{0}$ is the scale
height of the distribution, and typically $\beta \approx
3/4$. Assuming the bubble was initially inflated at the cluster
centre, and substituting for both $r_{\rm b}$ and $B$, the heating
rate can be re-written as
\begin{eqnarray}\label{eq:14a}
\dot{E} \approx 2 \times 10^{37}\bigg(\frac{r_{\rm b,0}}{5\,{\rm
    kpc}}\bigg)^{2} \bigg(\frac{P_{0}}{10^{-10}\,{\rm
    erg\,cm^{-3}}}\bigg)^{3/2}\\ \nonumber \times
\bigg(\frac{T}{10^{8}\,{\rm
    K}}\bigg)^{-2}\bigg[1+\bigg(\frac{z}{z_0}\bigg)^{2}\bigg]^{-\beta}\,{\rm
  erg\,s^{-1}}.
\end{eqnarray}
The thickness of a shell around the bubble that will be heated by
Ohmic dissipation can be estimated from $\dot{E}_{\rm e} = 4 \pi
r_{\rm b}^{2} \Delta r n^{2}\Lambda(T)$, where $n^{2}\Lambda(T)$ are
the radiative losses due to thermal bremsstrahlung. Using similar
quantities to those above, it can be shown that $\Delta r \sim
10^{-5}-10^{-2}\,$kpc, which will not be observable.

Therefore, Ohmic dissipation should be insignificant compared to the
total radiative ICM cooling rate due to thermal bremsstrahlung and is
unlikely to have a noticable effect even in a thin shell around the
bubble. Furthermore, even if the conductivity was drastically reduced,
the small volume occupied by the induced current means that the total
heating rate will always be negligible compared to the radiative
losses of the ICM.

\subsection{Total current}

The current generated by the Lorentz force flows in a thin layer
across the surface of the bubble. Assuming, again, that the bubble is
spherical with a radius $r_{\rm b}$, the total available area over
which the current can flow is $ 4 \pi [(r_{\rm b} + r_{\rm g})^{2} -
  r_{\rm b}^{2}] \approx 8 \pi r_{\rm b} r_{\rm g}$. Then, using
equation (\ref{eq:9}), the total current flowing around a typical
bubble is
\begin{equation}\label{eq:12}
I \approx 8\pi r_{\rm b}\displaystyle\sum_{i=p,e} J_{\rm i}r_{\rm g,i}
= \frac{16 \pi}{\mu_0}r_{\rm b}B \approx 6 \times
10^{18}\bigg(\frac{r_{\rm b}}{5\,{\rm
    kpc}}\bigg)\bigg(\frac{B}{10\,{\rm \mu G}}\bigg)\,{\rm A},
\end{equation}
and is independent of the gyration radius.

The magnitude of the current given in equation (\ref{eq:12}) is
somewhat uncertain because although the current loop probably closes
somewhere at the rear of the bubble, it is impossible to say precisely
where. Because of this uncertainty, the current may only flow over a
fraction of the bubble surface, thereby reducing the total
current. However, the reduction is unlikely to be more than an order
of magnitude.

Nevertheless, the value of the current in equation (\ref{eq:12}) is
almost identical to the values expected in current-dominated jets
\citep[e.g.][]{nakamura,diehl28}. In fact, this is not a coincidence -
the electric current and bubble radius associated with a magnetic
field in pressure equilibrium with its surroundings are related by
Amp\`{e}re's law. However, the values arise from different starting
points. For the current-dominated jet, described by \cite{nakamura},
the constant current generates a magnetic field; the bubble radius can
then be calculated assuming the field is in pressure equilibrium with
the ICM. In contrast, we have calculated the induced current for an
assumed magnetic field (in pressure equilibrium with its surrundings)
and bubble radius. The main difference is that the current estimated
in this way is not constant, and decreases as the bubble ascends
through the ICM.

\section{Discussion}

In the idealised model of the interaction between the ICM and the
AGN-blown bubble, described above, there is no mass or energy transfer
across the interaction boundary (as well no drag force exerted on the
bubble). If this is true, then it is not clear there should be any
thermal material in the bubble at all, while recent observations
indicate that thermal material may occupy up to 50\% of the volume of
some bubbles in the Perseus cluster \citep[][]{sand}. There are three
main possibilities which can explain this discrepancy:

1) Thermal material from the ICM was entrained into the outflow that
inflated the bubble. It may also be important to note that if a
supersonic outflow entrains material, it will be decelerated so that
its Mach number tends towards unity \citep[][]{hart86}. This effect
may explain prevalence of shocks in the ICM with Mach numbers close to
unity \citep[e.g.][]{form8707,grah}, as well as the origin of thermal
material in the bubble.

2) The second possibility is that charged particles diffuse into the
bubble. Due to the initial expansion and the effect of magnetic
draping, the magnetic fields will be primarily tangential to the
bubble surface. Therefore, in order to enter the bubble, particles
must diffuse across the field lines. Unlike diffusion parallel to the
magnetic field lines, which can occur rapidly, diffusion across field
lines is significantly reduced by many orders of magnitude
\citep[c.f.][]{enss03}. In fact, the diffusion coefficients
perpendicular to the field lines derived by \cite{enss03} are more
than 10 orders of magnitude less than those employed in numerical
simulations by \cite{rusz07,mat08,mat09}. We also note that, since the
bubble ascends at a velocity, $w$, with respect to the material
entering it, the transfer of mass is accompanied by a retarding force,
$-\dot{M}w$, where $\dot{M}$ is the rate at which mass enters the
bubble. Depending on the magnitude of the mass flux, the retarding
force may play an important role in governing the terminal velocity of
the bubble. The magnitude of the diffusive mass flux, therefore, has
potentially important consequences for interpreting observations of
bubbles.

3) Magnetic reconnection occurs in a medium with non-zero resistivity
(e.g. the ICM) when oppositely directed magnetic fields are brought
close together \citep[][]{boden}. Importantly, when the magnetic
fields of the ICM and bubble become interconnected, the field attains
a finite component, $B_{\rm n}$, normal to the bubble surface,
allowing plasma to flow across the bubble/ICM boundary
\citep[e.g.][]{pasch}. Transfering mass in this way would deliver a
retardation to the bubble, and may also generate convective motions
within the bubble that could affect its radio emission.

Regarding point 1), material may be entrained into an outflow due to
the growth of Kelvin-Helmholtz (K-H) instabilities at the boundary
between the outflow and ambient medium \citep[e.g.][]{dey06}. These
instabilities form as a result of the velocity shear between the two
fluids and have been shown to occur in magnetohydrodynamic (MHD) flows
\citep[][]{ryuj} as well as purely hydrodynamic flows. If the masses
entrained by hydrodynamic outflows, $10^{7}-10^{9}\,{\rm M_{\odot}}$
\citep[e.g.][]{dey86,reynolds}, are applicable to real AGN outflows it
is possible that this mechanism can explain the thermal content of
AGN-blown bubbles. However, recent work \citep[][]{nak} suggests that
magnetic tower jets do not develop K-H instabilities. Thus, if AGN
outflows follow this description, the thermal content of bubbles must
arise from either cross-field diffusion or magnetic reconnection,
discussed below.

The diffusive mass flux, per unit area, can be written $F = D \nabla
\rho$, where $D$ is the diffusion coefficient and $\nabla \rho$ is the
density gradient. The total mass flux into the bubble is then obtained
by integrating $F$ over the surface of the bubble, which, for a
spherical bubble gives $\dot{M}_{\rm diff} = 4\pi r_{\rm
  b}^{2}F$. Furthermore, assuming that the bubble is initially devoid
of thermal content and the bubble/ICM boundary has a thickness of
roughly one proton gyration radius, the density gradient, $\nabla
\rho$, can be no greater than $\rho/r_{\rm g,p}$. As a result, the
rate at which mass diffuses into a spherical bubble cannot exceed
$4\pi r_{\rm b}^{2}D\rho/r_{\rm g,p}$.

A convenient way of expressing the importance of diffusion is the
`refilling' timescale over which the diffusive influx of material
becomes comparable to the mass of ICM material displaced by the
bubble, $\tau_{\rm diff} \equiv M_{\rm dis}/\dot{M}_{\rm diff}$. Using
$M_{\rm dis} = (4/3)\pi r_{\rm b}^{3}\rho $ therefore, gives an upper
limit on the refilling timescale due to diffusion
\begin{eqnarray}\label{eq:time}
\tau_{\rm diff} \approx \frac{1}{3}\frac{r_{\rm b}r_{\rm g,p}}{D}
\approx 0.1 \bigg(\frac{D}{10^{16}\,{\rm
    cm^{2}\,s^{-1}}}\bigg)^{-1}\bigg(\frac{r_{\rm b}}{5\,{\rm
    kpc}}\bigg)\\ \nonumber \times \bigg(\frac{B}{10\,{\rm
    \mu\,G}}\bigg)^{-1}\bigg(\frac{T}{10^{8}\,{\rm
    K}}\bigg)^{1/2}\,{\rm Gyr}.
\end{eqnarray}
Equation (\ref{eq:time}) shows that a typical cross-field diffusion
coefficient of $D \sim 10^{16}\,{\rm cm^{2}\,s^{-1}}$ is required to
transfer a significant quantity of ICM material over a bubble lifetime
of 100 Myr.

The diffusion coefficient must fall within the range $ D_{\rm clas}
\le D \le D_{\rm B}$, where $D_{\rm clas}$ and $D_{\rm B}$ are the
classical and Bohm diffusion coefficients, respectively
\citep[][]{ikhsanov}. Classically, the diffusion of charges across
magnetic field lines arises from non-zero plasma resistivity
\citep[e.g.][]{spitzer,gold}, that is, it scales with the collision
rate of electrons with protons, $\nu_{\rm e,p}$, and is roughly
$D_{\rm clas} \approx \nu_{\rm e,p} r_{\rm g,e}^{2}$
\citep[e.g.][]{gold}, where $r_{\rm g,e}$ is the electron gyration
radius. For typical cluster values, $D_{\rm clas} \sim 10 {\rm
  cm^{2}\,s^{-1}}$ and, therefore, will not lead to significant mass
transfer from the ICM to the bubble on a realistic timescale.

The maximum diffusive mass transfer rate occurs if diffusion is
governed by drift-dissipative instabilities - Bohm diffusion
\citep[e.g.][]{ikhsanov,enss03}, for which $D_{\rm B} \approx k_{\rm
  B}T/(16 e B) \approx 5 \times 10^{15}(T/10^{8}\,{\rm
  K})(B/10\mu\,{\rm G})^{-1}\,{\rm cm^{2}\,s^{-1}}$. Therefore, mass
transfer due to Bohm diffusion would be able fill a bubble on a
timescale of 100 Myr. Despite being an extreme upper limit, the Bohm
limit does, nevertheless, provide an estimate of the minimum possible
timescale a bubble can survive before diffusion-driven mass transfer
completely refills it.

Perhaps more realistically, \cite{enss03} calculated diffusion
coefficients for cosmic rays crossing a turbulent magnetic field to be
$\sim 10^{15}\,{\rm cm^{2}\,s^{-1}}$, using $B = 10\,{\mu G}$. For
thermal particles with an energy of $\sim 10\,{\rm keV}$, the
diffusion coefficient is reduced to $D \sim 10^{11}\,{\rm
  cm^{2}\,s^{-1}}$. In this case, the time taken to completely refill
the bubble would be $10^{12}\,{\rm yr}$. Therefore, unless diffusion
proceeds at the Bohm rate, it is unlikely to contribute a significant
propertion of the thermal content of bubbles.

If magnetic reconnection occurs, the plasma transfer rate is
determined by the dimensionless reconnection rate commonly defined as
$\zeta = B_{\rm n}/B$. However, even for the Earth's magnetosphere it
is difficult to accurately determine $B_{\rm n}$, though typical
values indicate $\zeta \sim 0.1$
\citep[][]{elsner,pasch,ikhsanov}. The mass flux entering the bubble
is then given by $\dot{M}_{\rm rec} = \pi r_{\rm b}^{2}\rho w \zeta$,
where $w$ is the ascent velocity of the bubble \citep[][]{pasch}. The
timescale for refilling a bubble as a consequence of magnetic
reconnection is then
\begin{eqnarray}\label{eq:timerec}
\tau_{\rm rec} \equiv \frac{M_{\rm dis}}{\dot{M}_{\rm rec}} =
\frac{4}{3}\frac{r_{\rm b}}{w \zeta} \approx 0.7 \bigg(\frac{r_{\rm
    b}}{5\,{\rm kpc}}\bigg) \bigg(\frac{w}{10^{7}\,{\rm
    cm\,s^{-1}}}\bigg)^{-1}\\ \nonumber \times
\bigg(\frac{\zeta}{0.1}\bigg)^{-1}\,{\rm Gyr}.
\end{eqnarray}
According to equation (\ref{eq:timerec}), it seems physically possible
that magnetic reconnection could explain the observations of
\cite{sand}. Furthermore, the transfer of mass would also be
accompanied by heating as the energy density of the magnetic field is
transferred to the plasma. However, the significance of this channel
of mass transfer depends crucially on the reconnection rate, $\zeta$,
which is highly uncertain.

\section{Summary}

The aim of this article has been to describe one of the main processes
which prevents thermal material from the ICM entering AGN-blown
bubbles. As a direct consequence, it is possible to highlight the
mechanisms which must occur if a bubble is to display thermal
content. The main findings are summarised below:

i) The Lorentz force causes a charged particle entering the bubble to
execute a half-orbit so that it is re-directed back into the ICM and
excluded from the bubble. The transverse motions of the deflected
charges generate electric currents along the bubble surface which
prevent the magnetic field from entering the ICM
\citep[e.g.][]{chap,funaki}. Therefore, once back in the ICM, a charge
experiences no further deflection. The charges penetrate roughly one
proton gyration radius before being directed back into the ICM. This
exclusion is different to the stabilising effect of magnetic fields
\citep[e.g.][]{chand,deyoung03,instab105,dursi07} which, strictly
speaking, only explains why the bubbles do not fragment into smaller
structures due to Rayleigh-Taylor instabilities. The combination of
both effects provides an effective surface tension which, in the
idealised model, prevents material from the ICM entering the bubble.

ii) Ohmic dissipation of the induced current heats the ICM adjacent to
the bubble at a rate $\sim 10^{35}-10^{37}\,{\rm erg\/s^{-1}}$; this
is insignificant compared to the radiative losses of the ICM. The
total induced current on the bubble surface has a typical magnitude of
$I \sim 10^{18}\,{\rm A}$, which is similar to that required to
produce bubbles from current-dominated jets
\citep[][]{diehl28,nakamura}.

iii) \cite{sand} recently presented observations indicating that
thermal material may occupy up to 50\% of the volume of some bubbles
in the Perseus cluster. Given the effect of the Lorentz force, the
most likely explanations for the observed thermal content are that:
mass is entrained into the outflow that inflated the bubble, charged
particles diffuse rapidly across the bubble/ICM boundary, or magnetic
reconnection on the bubble surface allows material to enter the
bubble. We find that, unless the AGN outflow behaves as a magnetic
tower jet, entrainment into the outflow can supply sufficient thermal
material to explain the results of \cite{sand}. However, for magnetic
tower jets, the outflow entrainment rate is expected to be
insignificant \citep[][]{nak}. The cross-field diffusion coefficient
required for thermal material to fill a typical bubble is $\sim
10^{16}\,{\rm cm^{2}\,s^{-1}}$. This value is anomalously high
compared to predictions from turbulent diffusion models
\citep[][]{enss03}, but is comparable to the Bohm diffusion
coefficient. Given that the Bohm limit is an extreme upper limit, it
seems unlikely that cross-field diffusion contributes significantly to
a bubble's thermal content. Finally, the mass transfer rate due to
magnetic reconnection is somewhat uncertain, but significant for
plausible reconnection rates. We conclude, based on current models,
that entrainment into the outflow and mass transfer due to magnetic
reconnection are probably the most significant sources of thermal
content in AGN-blown bubbles.

\section{Acknowledgements}
The author would like to CITA for funding through a National
Fellowship, informative discussions with Jim Hinton, and the referee
for helpful comments that improved this work.

\bibliography{database} \bibliographystyle{mn2e}

\label{lastpage}

\end{document}